\documentclass[12pt]{article}

\usepackage{scicite}
\usepackage{caption}
\usepackage{times}
\usepackage{afterpage}
\usepackage{floatpag}
\usepackage{graphicx} 
\usepackage{color}    

\newenvironment{sciabstract}{%
\begin{quote} \bf}
{\end{quote}}

\title{Laser-assisted Fano resonance: attosecond quantum control and dynamical imaging} 

\author
{Meng Han$^{1,\ast,\dagger}$, Hao Liang$^{2,\ast,\dagger}$, Jia-bao Ji$^{1}$, Leung Chung Sum$^{1}$, \\Kiyoshi Ueda$^{1,3}$, Jan Michael Rost$^{2}$, Hans Jakob Wörner$^{1}$
\\
\normalsize{$^{1}$Laboratorium für Physikalische Chemie, ETH Zürich, Zürich, 8093, Switzerland}\\
\normalsize{$^{2}$Max Planck Institute for Physics of Complex Systems, Nöthnitzer Straße 38, 01187, Germany}\\
\normalsize{$^{3}$Department of Chemistry, Tohoku University, Sendai, 980-8578, Japan}\\
\normalsize{$^\ast$These authors contributed equally to this work}\\
\normalsize{$^\dagger$E-mails: menhan@ethz.ch; liangh@pks.mpg.de}\\
}

\date{\today}

\begin{document} 

\baselineskip20pt
\maketitle 

\begin{sciabstract}
A Fano resonance arises from the pathway interference between discrete and continuum states, playing a fundamental role in many branches of physics, chemistry and material science. Here, we introduce the concept of a laser-assisted Fano resonance, created from two interferometric pathways that are coupled together by an additional laser field, which introduces a controllable phase delay between them and results in a generalized Fano lineshape that can be actively controlled on the {\it attosecond} time scale. Based on our experimental results of unprecedented resolution, we dynamically image a resonant electron wave packet during its evolution directly in the time domain, extracting both the amplitude and the phase, which allows for the measurement of the {\it resonant} photoionization time delay. Ab-initio calculations and simulations employing a physically transparent two-level model agree with our experimental results, laying the groundwork for extending our concepts into attosecond quantum control of complex systems.      
\end{sciabstract}

\clearpage
We live in a world ruled by resonances, ranging from atoms over musical instruments to medical equipment. Spectroscopy is an essential tool for uncovering resonances in light-matter interactions. The observed spectral line positions reveal resonant energies of the excited quantum states, whereas the lineshapes are determined by how the matter relaxes after light is resonantly absorbed. For example, when a discrete state is resonantly excited by a light field, the corresponding absorption spectrum will show a symmetric Lorentzian lineshape. If the discrete state is coupled with a continuum, the pathway interference between them will result in an asymmetric Fano lineshape \cite{fano1935sullo,fano1961effects}. As a central paradigm, Fano resonances widely exist in many branches of science \cite{rost1997resonance,fan2010self,limonov2017fano,mazumdar2006efimov}, serving as a bridge towards quantum optical control of electron dynamics, such as electromagnetically induced transparency (EIT) \cite{alzetta1997induced}, Borrmann \cite{novikov2017borrmann} and Kerker \cite{rybin2013fano,poshakinskiy2019optomechanical} effects.

The birth and development of attosecond light sources \cite{Hentschel2001,Paul2001} and attosecond metrologies \cite{itatani2002attosecond,schultze2010delay,dahlstrom2013theory} have enabled real-time observation of ultrafast electron dynamics inside atoms \cite{ott2013lorentz,kaldun2016observing,gruson2016attosecond}, molecules \cite{Huppert2016,gagnon2007soft}, solids \cite{cavalieri2007attosecond,Ghimire2011} and liquids \cite{Luu2018,Jordan2020}. Consequently, the quest for quantum control of the electron dynamics on the attosecond time scale becomes a central topic in ultrafast science \cite{lin2013controlling,kraus15b}. Although quantum control in infrared (IR) and visible frequency regimes has already been accessible based on the dense energy levels of cold alkali-metal atoms \cite{chu2002cold} or the tunable plasmonic resonance of nanostructures \cite{limonov2017fano}, extending such manipulations down to the extreme-ultraviolet (XUV) regime is still quite challenging owing to both technical and methodological limitations. Recent breakthroughs \cite{goulielmakis2007attosecond,bandrauk2002attosecond} have demonstrated the control of attosecond dynamics in a classical way, relying on the electron classical motion driven by optical fields. However, attosecond quantum control of the electron dynamics is still in its infancy.

\begin{figure}[htbp]
\centering
\includegraphics[width=\linewidth]{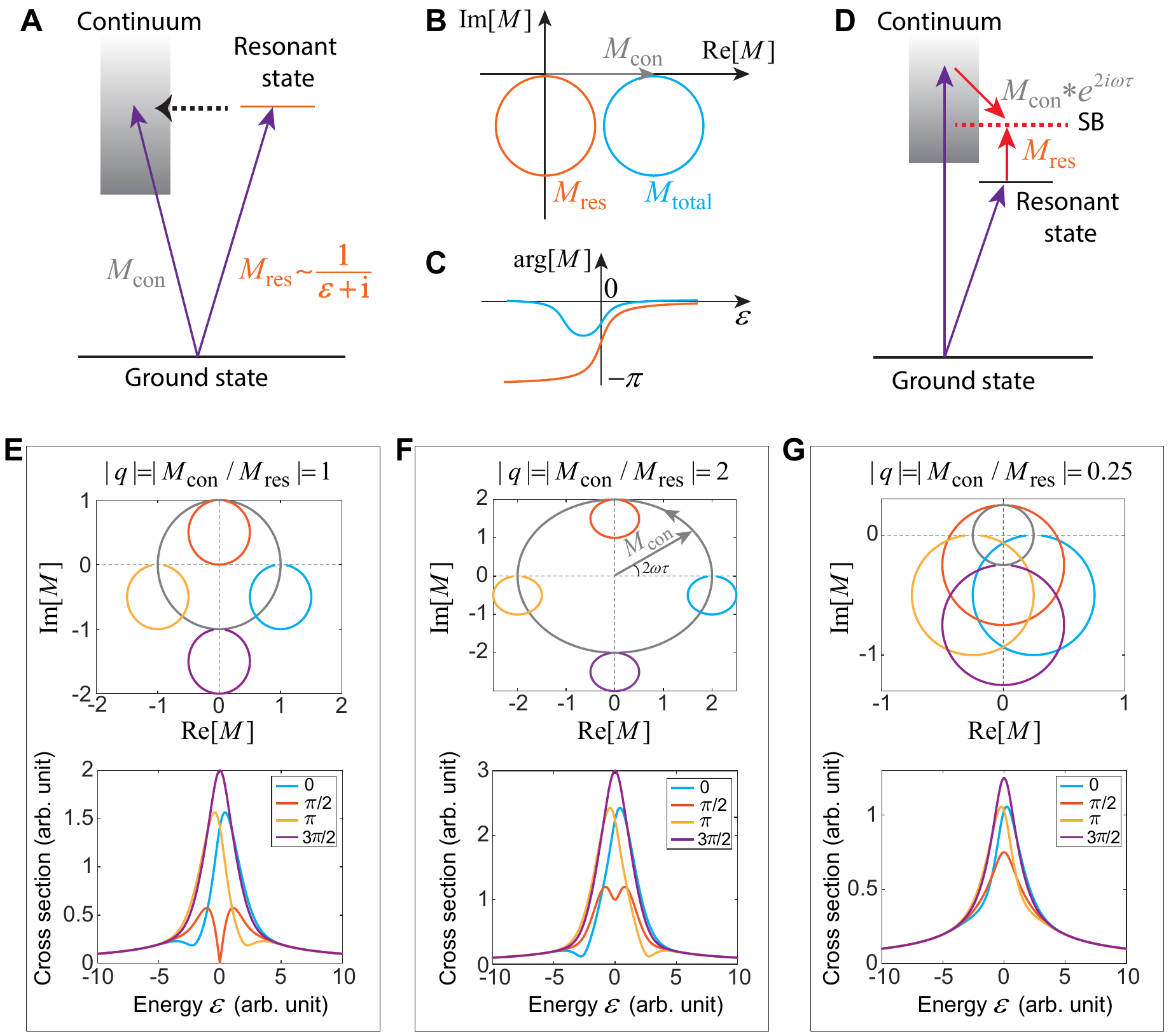}
\caption{\textbf{Principle of laser-assisted Fano resonance and its quantum control.} (\textbf{A}) Schematic diagram of a standard Fano resonance, which is given by the interference of a continuum pathway $M_{\rm{con}}$ and a resonant pathway $M_{\rm{res}}$. The resonant state is embedded in the continuum. (\textbf{B}) Complex-plane representations of $M_{\rm{con}}$, $M_{\rm{res}}$ and their sum $M_{\rm{total}}$. The resonant transition amplitude is shifted to become asymmetric with respect to the imaginary axis after the coherent superposition with a real-valued continuum transition amplitude. (\textbf{C}) Phases of $M_{\rm{res}}$ (orange curve) and $M_{\rm{total}}$ (blue curve) as a function of the reduced energy $\varepsilon$, where $\varepsilon = 0$ stands for the resonance energy. Note that the phase extreme of $M_{\rm{total}}$ is offset with respect to the resonant position. (\textbf{D}) Schematic diagram of laser-assisted Fano resonance, where the resonant state lies below the continuum. The resonant and continuum pathways are respectively triggered by two adjacent odd harmonics and their recombination is assisted by an IR laser field. Taking $M_{\rm{res}}$ as a reference, $M_{\rm{con}}$ will carry a phase factor $e^{2i\omega\tau}$, corresponding to a circle centered on the origin of the complex plane. (\textbf{E-G}) Complex-plane representations (upper panels) and lineshapes (bottom panels) for laser-assisted Fano resonance when the channel amplitude ratio is 1.0 (E), 2 (F) and 0.25 (G). The XUV-IR delay $\tau$ decides where the Lorentz circle rotates on the concentric circle and therefore controls the lineshape. In the calculations, we adopt $M_{\rm{res}}=\frac{1}{\varepsilon+\rm{i}}$ and $M_{\rm{con}}= qe^{2i\omega\tau}\cdot e^{-\varepsilon^2/4}$ assuming a Gaussian-shaped light spectrum, where $q$ is the channel-amplitude ratio shown on the top of the panels. }
\label{fig:figure1}
\end{figure}

Here, we show how to quantum engineer the Fano lineshape through photoelectron spectroscopy by introducing a new variant of the Fano resonance, i.e. the laser-assisted Fano resonance. Previous time-resolved studies have been limited to the well-known ``natural" Fano resonances arising from the doubly excited states of helium \cite{ott2013lorentz,kaldun2016observing,gruson2016attosecond}. In the standard Fano resonance, the resonant state is a doubly excited state, lying higher than the first ionization threshold of the system, as shown in Fig. \ref{fig:figure1}A. Therefore, an XUV pulse can eject the electron directly into the continuum or excite it into the resonant doubly excited state. The latter can then autoionize and couple to the continuum wave packet, whereby their interference gives rise to the Fano lineshape. 

The relationship between the interference and the asymmetric lineshape becomes clearer when one uses the complex-plane representation, as shown in Fig. \ref{fig:figure1}B, where $M_{\rm{con}}$, $M_{\rm{res}}$ and $M_{\rm{total}}$ stand for complex amplitudes of the continuum pathway, the resonant pathway and their coherent sum, respectively. The resonant transition amplitude can be described by a Breit-Wigner amplitude \cite{gruson2016attosecond} $\sim \frac{1}{\varepsilon+\rm{i}}$ whose absolute square corresponds to a Lorentzian lineshape, where $\varepsilon = (E-E_r)/(\hbar\Gamma/2)$ denotes the reduced energy containing $E_r$ and $\Gamma$ as the position and width of the resonance, respectively, and $\hbar$ is the reduced Planck constant. The resonant transition amplitude (orange circle in Fig. \ref{fig:figure1}B), which is originally symmetric with respect to the imaginary axis, will be shifted to become asymmetric with respect to the imaginary axis (blue circle in Fig. \ref{fig:figure1}B) after the coherent superposition with a real-valued continuum transition amplitude. The position of the transition amplitude in the complex plane determines the argument (i.e. phase) as well as the modulus (i.e. cross section). The phase distributions of $M_{\rm{res}}$ and $M_{\rm{total}}$ as a function of energy are depicted in Fig. \ref{fig:figure1}C, where the Lorentzian curve has a $\pi$-phase jump at the resonant position and the phase minimum position of the Fano curve deviates from the resonant position. This energy offset for the phase is related to the asymmetry parameter $q$ (real number) of its lineshape, which is defined as the channel intensity ratio \cite{fano1961effects}, and it can be influenced by the Stark shift of the resonant state in a strong IR field, as illustrated in previous studies \cite{ott2013lorentz,ott2019strong}. However, this line-shape control relies on the action of the pulse envelope and is thus limited to the femtosecond time scale.

Here, we introduce the concept of the laser-assisted Fano resonance, as illustrated in Fig. \ref{fig:figure1}D. We separate the two interferometric arms of the Fano resonance and introduce a controllable phase delay between them before the recombination. The resonant and continuum pathways can be respectively triggered by two adjacent odd harmonics in an attosecond pulse train, and their recombination occurs at the intermediate sideband (SB) through the assistance of a fundamental IR field. By coupling with the IR field, the two pathways acquire phase factors of $e^{i\omega\tau}$ and $e^{-i\omega\tau}$, respectively, where $\omega$ is the IR frequency and $\tau$ is the XUV-IR time delay. By taking $M_{\rm{res}}$ as the reference, $M_{\rm{con}}$ will carry a phase factor of $e^{2i\omega\tau}$, corresponding to a concentric circle in the complex plane. Therefore, the coherent sum of $M_{\rm{res}}$ and $M_{\rm{con}}e^{2i\omega\tau}$ corresponds to the rotation of the resonant Lorentzian circle on the concentric circle of the continuum pathway, and the azimuth angle of the rotation is their phase difference $2\omega\tau$, which is experimentally controllable on the attosecond time scale. Therefore, in our case the $q$ parameter, $q = |M_{\rm{con}}/M_{\rm{res}}|e^{2i\omega\tau}$, is a complex number, which allows to rotate the transition amplitude in the whole complex plane and thus covers the concept of the ``natural" Fano resonance. 

In Fig. \ref{fig:figure1}E, we show the complex-plane representations and the lineshapes of the total transition amplitude $M_{\rm{total}}$ at the phase delays of 0, $\pi/2$, $\pi$, and $3\pi/2$. For the cases of $\pi/2$ and $3\pi/2$, which correspond to symmetric double-peak and single-peak lineshapes, respectively, the total amplitudes in the complex plane are both symmetric with respect to the imaginary axis. At the delays of 0 and $\pi$, $M_{\rm{total}}$ rotates to the positions which are the farthest away from the imaginary axis, and thus the asymmetry of the corresponding lineshapes is the largest, with the opposite directions for the two delays. Note that not only the intensity ratio $|M_{\rm{con}}|/|M_{\rm{res}}|$, i.e., the traditional Fano $q$ parameter, but also phase delay between the two arms determine the lineshape in our case. The intensity ratio is visualized as the relative size of the two circles in the complex plane. In Fig. \ref{fig:figure1}F-G, we display the lineshapes at $|M_{\rm{con}}|/|M_{\rm{res}}|$ of 2 and 0.25, respectively. When $|M_{\rm{res}}|$ is larger than $|M_{\rm{con}}|$, the two peaks at the delay of $\pi/2$ might be merged together and thus the lineshape (red curve in Fig. 1G) goes back to a symmetric single peak. Furthermore, both the peak-position shift between the delays of 0 and $\pi$ and the peak-amplitude contrast between the delays of $\pi/2$ and $3\pi/2$ decrease with the decrease in the ratio $|M_{\rm{con}}|/|M_{\rm{res}}|$. In reality, the intensity ratio between the two pathways can be controlled by the Stark-shift-induced detuning of the resonance in the femtosecond time scale, as we show later in our experiment.

Experimentally, the XUV attosecond pulse train is produced from high-order harmonic generation of argon atoms driven by an intense IR pulse accurately characterized by the transient-grating FROG technique. The central wavelength is 799 nm and the pulse duration (FWHM) is 31 fs. The XUV pulse duration (FWHM) is characterized to be 12 fs from the cross-correlation signal of normal sidebands. We choose the singly excited 1s3p state of helium atoms as the resonant state to demonstrate the concept of laser-assisted Fano resonance. This resonant state is naturally covered by the harmonic 15 with the bandwidth of $\sim$0.2 eV and is identified in an additional experiment of the long-range XUV-IR scan [see Supplementary Material (SM)]. The harmonic 17 can directly ionize the helium atoms, serving as the interferometric arm of the continuum pathway. The sideband 16 is formed by the interference of the two pathways, i.e. the signal of laser-assisted Fano resonance. Our experimental approach builds on the across-threshold RABBIT scheme \cite{swoboda2010phase,neorivcic2022resonant,autuori2022anisotropic,drescher2022phase,kheifets2021rabbitt,villeneuve2017coherent}, but critically hinges on two aspects: (i) we have dramatically enhanced the energy resolution to 10 meV around zero kinetic energy using a COLTRIMS detector \cite{dorner00a,ullrich03a}, and (ii) we have resolved the XUV-IR delay in a very long range of 50 fs with the attosecond stability. The former allows one to directly observe the attosecond change of the lineshapes, and the latter reveals that the RABBIT phase is locally dependent on the XUV-IR delay, adding a fundamentally new aspect compared to previous studies \cite{swoboda2010phase,neorivcic2022resonant,autuori2022anisotropic,drescher2022phase,kheifets2021rabbitt,villeneuve2017coherent}.

\begin{figure}[htbp]
\centering
\includegraphics[width=\linewidth]{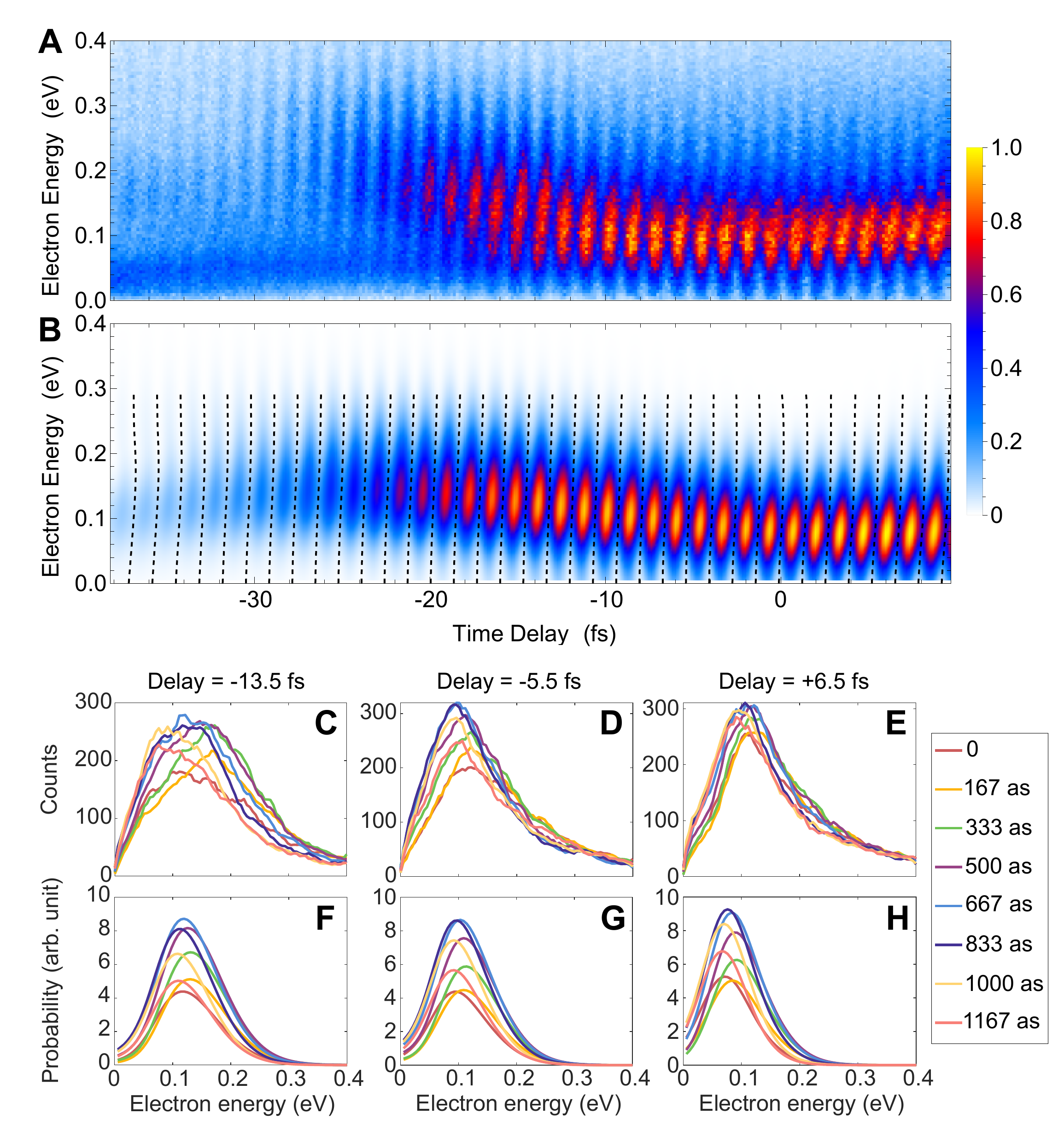}
\caption{\textbf{Temporal control of Fano lineshapes.} (\textbf{A-B}) Delay-resolved photoelectron energy spectra of the sideband 16 from measurement (A) and TDSE calculation (B). In (B) the dashed lines are the interference minima predicted by the two-level model, which shows excellent agreement with the TDSE calculation. (\textbf{C-H}) Attosecond-resolved photoelectron lineshapes at different femtosecond delays labeled on the top of the panels from experiment (C,D,E) and TDSE calculation (F,G,H).}
\label{fig:figure2}
\end{figure}

In Figs. \ref{fig:figure2}A-B, we show the delay-resolved photoelectron energy spectra of the sideband 16 from the measurement and the ab initio solution of time-dependent Schr\"odinger equation (TDSE), respectively. In our TDSE simulations, we used the modified Tong-Lin potential \cite{tong05a}, and the calculated resonant energy of 1s$^2$-1s3p transition agrees very well with the NIST value ($0.84843\,$a.u.) down to the fifth significant digit. In the simulation, the XUV and IR pulses are both of Gaussian shape without any chirps, and the light intensities and durations are adopted from the experimental values. The photoelectron final momentum or energy is obtained by projecting the wavefunction onto the continuum eigenstates of the model potential without any approximations (see SM for the calculation details). The TDSE simulation reproduces most of features we measured, including both femtosecond and attosecond structures. In our definition, the negative time delay means that the IR pulse is behind the XUV pulse. Around the delay of zero, the photoelectron energy is the lowest, owing to the largest Stark shift (by the ponderomotive energy $U_p$) of the ionization potential ($I_p$) caused by the peak of the IR envelope. When the IR pulse gradually comes after the XUV pulse, the electron energy goes up due to the decrease of the Stark shift until around -25~fs, where the two pulses only overlap by a small part and thus the continuum pathway becomes very weak. Since the resonant pathway has a much weaker Stark-shift effect than the continuum pathway, the electrons from the two pathways will gradually mismatch in energy when delaying the IR pulse until leaving a long tail of the resonant pathway in the lower-energy regime. As for the attosecond structure, both measurement and TDSE simulation reveal that the RABBIT fringes are tilted and the slope locally depends on the time delay. In Figs. \ref{fig:figure2}C-E, we display the line cuts with attosecond resolution at the time delays of around -13.5~fs, -5.5~fs and 6.5~fs, respectively, which agree well with the corresponding TDSE results shown in Figs. \ref{fig:figure2}F-H. The photoelectron lineshapes varied from a low-amplitude symmetric one, over a rightward biased one and then a high-amplitude symmetric one, to a leftward biased one. This behavior basically corresponds to the physical picture that $|M_{\rm{res}}|$ is larger than $|M_{\rm{con}}|$ (i.e. Fig. \ref{fig:figure1}G). As the time delay decreases from 6.5 fs to -13.5 fs, the left-right energy shift and the up-down yield contrast are both enhanced, indicating that the two pathways become closer in channel intensity. This is because delaying the IR pulse will increase the detuning of the resonant pathway and then its channel intensity decreases faster than the continuum pathway, making the two channel intensities more comparable. Therefore, the attosecond control of the lineshape corresponds to the two-circle rotation in the complex plane, while the femtosecond control relies on the manipulation of the relative size of the two circles. Both can be well understood within the physical picture of the laser-assisted Fano resonance, as shown in Fig. \ref{fig:figure1}. 

To capture the physical processes at play, we develop a transparent two-level model, including two discrete states (1s and 3p) and a continuum. The resonant-pathway amplitude is described by $M_{\rm{res}} = -\mathrm i \mu_{3p,E}\int e^{iEt}F_{\rm{IR}}(t)c_{3p}(t)\,\mathrm dt$, and the continuum-pathway amplitude is written as $M_{\rm{con}} = -\mathrm i\pi\mu_{1s,E+\omega}\mu_{E+\omega,E}\int e^{iEt}F_{\rm{IR}}(t)F_{\rm{H17}}(t)c_{1s}(t)\,\mathrm dt$, where $\mu_{A,B}$ is the transition dipole from state $A$ to state $B$ ($A,B$ stand for the bound 1s and 3p states or the continuum states with energy of $E$ and $E+\omega$), $E$ is the sideband energy, $F_{\rm{IR}}$ and $F_{\rm{H17}}$ are the IR and harmonic 17 fields, respectively. The time-dependent probability amplitudes for the electron being pumped to the 3p state $c_{3p}(t)$ or staying in the 1s state $c_{1s}(t)$ are described with a two-level equation including  Stark shift and  ionization loss (see SM for more details). We calculate the energy-resolved interference minima between $M_{\rm{res}}$ and $M_{\rm{con}}$ based on the two-level model and plot them onto Fig. \ref{fig:figure2}B with dashed lines. The predictions of the two-level model agree excellently with the ab-initio TDSE result over the whole delay range. This finding gives rise to a counter-intuitive conclusion that the energy dependence of the Wigner phase and continuum-continuum (CC) transition phase \cite{dahlstrom2013theory} are negligible here even in this low-energy regime, since our two-level model did not include the Wigner or CC phase. In the SM, we show that the Wigner and CC phases cancel out in the across-threshold RABBIT scheme. Therefore, here the measured energy and delay dependence of the RABBIT fringes originates from the resonant dynamics.
This finding has important consequences because it defines an absolute time scale for the measurements of photoionization delays. Whereas previous RABBIT measurements could only determine {\it relative} photoionization delays between two simultaneously measured photoionization processes \cite{kluender11a,Huppert2016,isinger17a,Jordan2020}, our results show a pathway for determining the origin of time in RABBIT measurements, which could be further developed to measure {\it absolute} photoionization delays.

\begin{figure}
    \centering
    \includegraphics[width=\linewidth]{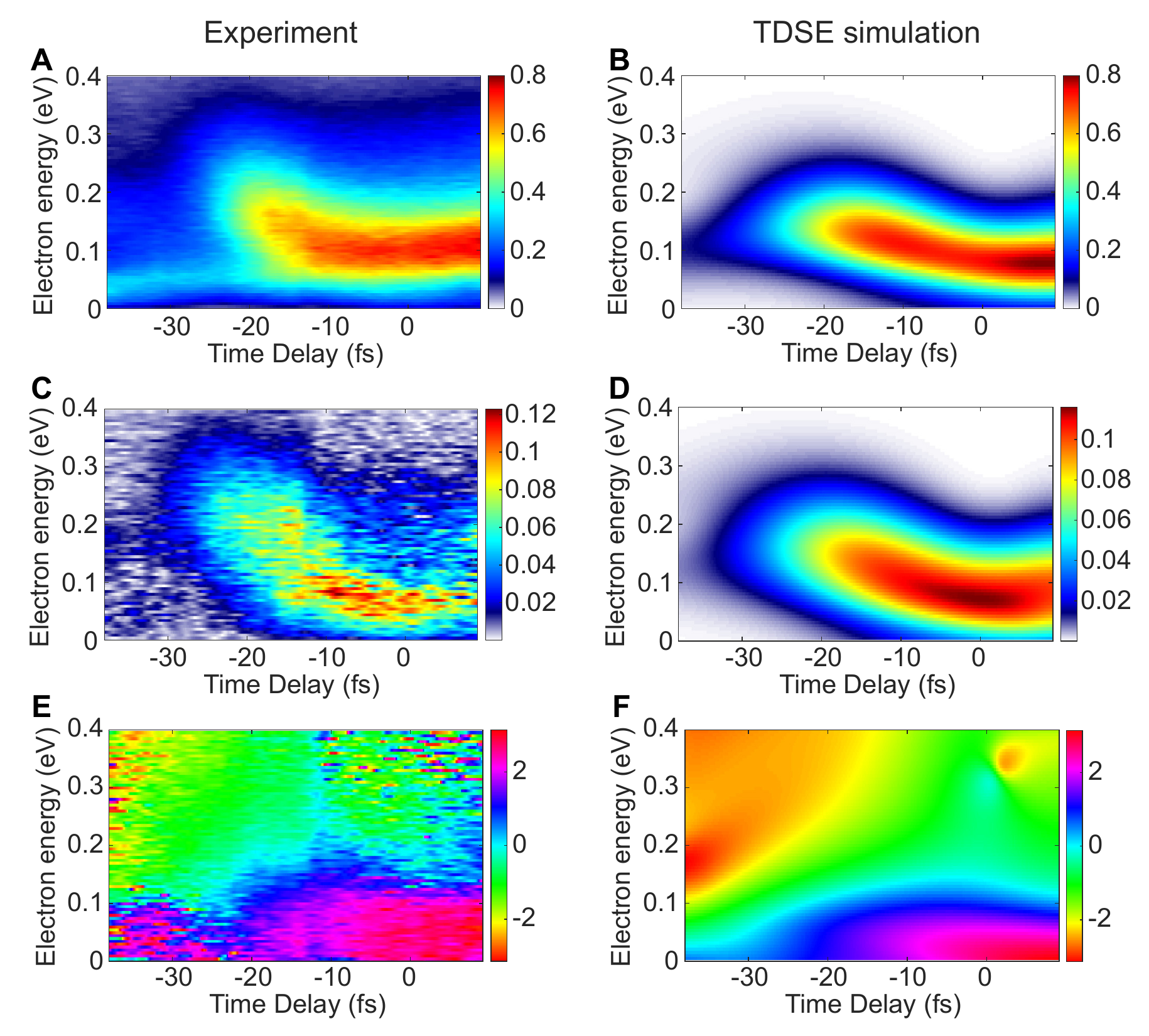}
    \caption{\textbf{Local fitting of RABBIT fringes.} The local fitting is performed by a cosine function of $a+2b\cos(2\omega\tau+\varphi)$ along the delay axis at each delay and energy. From top to bottom: $a,b$ and $\varphi$. Left and right panels are for experiment data and TDSE simulation, respectively.}
    \label{fig:figure3}
\end{figure}

To quantify the energy and delay dependence, we locally fit the interference fringes with the function $a+2b\cos(2\omega\tau+\varphi)$. In Fig. \ref{fig:figure3}, we show the fitting results of $a$, $b$ and $\varphi$ from the experimental data and the TDSE simulation, respectively. The background term $a$ corresponds to the sum of squares of the two wave packets, i.e. $|M_{\rm{con}}|^2 + |M_{\rm{res}}|^2$, while the fringe contrast $b$ corresponds to the product of their modulus, i.e. $|M_{\rm{con}}| \cdot |M_{\rm{res}}|$. Hence, the background term displays a low-energy tail if the two pulses are separated, as opposed to the contrast term. The parabolic shape of both $a$ and $b$ manifests the effect of the $U_p$-shift from the IR field in the continuum pathway. Note that here the IR light intensity is as low as in typical RABBIT experiments. Our experimental data reveals that the effect of the IR-pulse envelope is not negligible if the energy resolution is high enough. Finally, the RABBIT phase $\varphi$ corresponds to the phase difference between the two pathways, i.e. ${\arg}[M_{\rm{res}}] - {\arg}[M_{\rm{con}}]$, which mainly reflects the feature of the resonant pathway. As the time delay decreases, the energy position of the phase jump (blue area) gradually becomes smoother and shifts towards the low-energy regime.

\begin{figure}
    \centering
    \includegraphics[width=.8\textwidth]{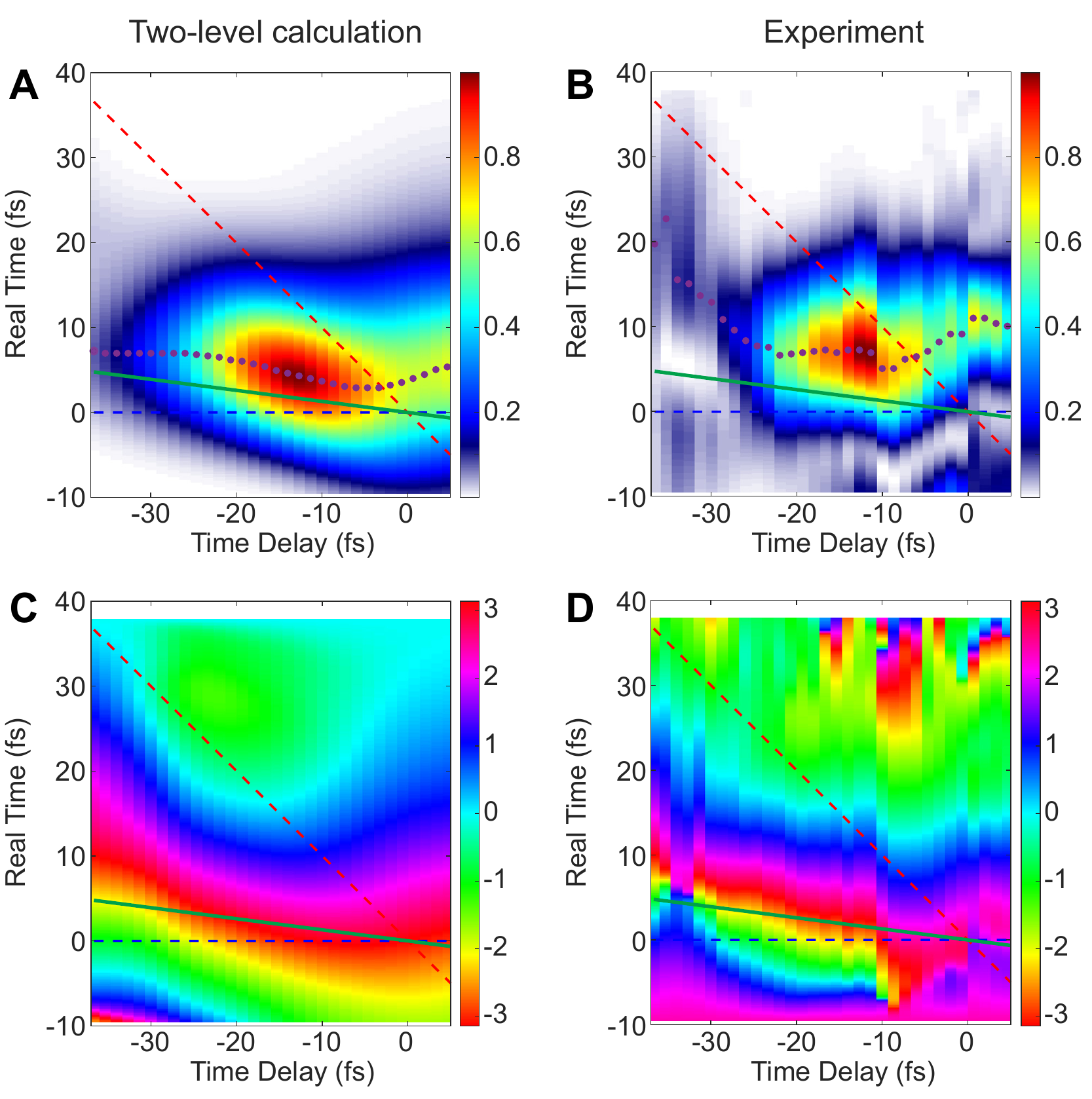}
    \caption{\textbf{Dynamical imaging of a resonant electron wave packet.} (\textbf{A-B}) Time- and delay-resolved amplitude ($|M_{\rm{res}}(t,\tau)|$) of the resonant wave packet from the two-level model (A) and measurement (B). The magenta dots plot the peak position of the wavepacket at each time delay. (\textbf{C-D}) The corresponding phase ($\arg[M_{\rm{res}}(t,\tau)]$) distribution from the two-level model (C) and measurement (D). In all panels, the blue and red dashed lines represent the peak positions of the XUV and IR pulses, respectively. The green solid line stands for the peak position of the product function of the XUV and IR pulses, corresponding to the center of the wave packet in the continuum pathway.}
    \label{fig:figure4}
\end{figure}

From the fitting results, we can extract the amplitude and the phase and thus dynamically image the resonant wave packet in the time domain. After subtracting the amplitude $|M_{\rm{con}}|$ and the phase $\arg[M_{\rm{con}}]$ of the continuum pathway, we obtain the complex amplitude $M_{\rm{res}}(E,\tau)$ of the resonant pathway in the energy domain. Using the time-frequency transformation, we finally obtain the wave packet $M_{\rm{res}}(t,\tau)$ in the time domain, as illustrated in Figs. \ref{fig:figure4}B-D (see SM for the details of data analysis). A comparison with the results directly calculated from the two-level model is shown in Fig. \ref{fig:figure4}A-C. In these plots, the horizontal axis stands for the XUV-IR time delay, while the vertical axis represents real time. The XUV pulse is centered at time zero (blue dashed line), and the IR pulse is located at positions where the real time equals to the XUV-IR time delay (red dashed diagonal line). We also plot the peak position of the product of two pulse envelopes (i.e. $I_{\rm{IR}}(t)I_{\rm{H17}}(t)$) as the solid green line, which corresponds to the peak position of the wave packet in the continuum pathway, i.e. the non-resonant two-photon wave packet. This solid green line is the reference that there is no any delays between absorbing XUV and IR photons. Both experimental and model results (see magenta dots in Figs. \ref{fig:figure4}(A-B)) reveal that the resonant wave packet peaks in the regime between the green and red lines, indicating that the resonant wave packet will stay at the resonant state for some time before absorbing the IR photon. The temporal separation between the peak positions of the resonant and non-resonant wave packets is the so-called resonant two-photon photoionization time delay \cite{su2014time}. 
Since the XUV pulse intensity is too weak to trigger Rabi flopping between the two levels, the population of the excited state generally accumulates within the XUV pulse. This accumulation leads to a delay in the instantaneous ionization from the resonant state.
Our finding validates a previous theoretical prediction \cite{su2014time} that the resonant two-photon time delay depends on the pulse envelope. The phase distributions exhibit increasing of the temporal dispersion of the electron wave packet when the IR pulse is delayed, which corresponds to the energy shift in the frequency domain. Note that we have varied pulse parameters in the two-level model and found the above discussed phenomena to be very general and robust. 

To conclude, we have quantum engineered a new variant of the Fano resonance, the laser-assisted Fano resonance, in which the photoelectron lineshape is controllable on both attosecond and femtosecond time scales. The different possibilities for control over lineshapes have been made transparent with a simple picture in the complex plane. Our unprecedented high-resolution experimental results reveal the envelope effects of XUV and IR pulses in the across-threshold RABBIT technique and show that the RABBIT phase locally depends on the two-pulse time delay. With the help of ab-initio calculations and a two-level model, we have clarified the role of the Wigner and CC delays in this across-threshold RABBIT case. We have further dynamically imaged a resonant electron wave packet and determined the resonant photoionization time delay. Since the resonant states slightly below the threshold can be regarded as bound states embedded in the continuum in the photon-dressed Floquet picture, the quantum manipulation based on Fano resonances, such as EIT \cite{alzetta1997induced}, should be easily accessible based on the singly excited states in attosecond science. For example, if one odd harmonic covers two singly excited states, this corresponds exactly to the case of a typical EIT system involving two Fano resonances \cite{limonov2017fano,lin2013controlling}. Hence, the photoelectron line due to three-channel interference will display a modulation between ``transparency'' and ``opaqueness'' that is controllable on the attosecond time scale. Therefore, our study paves the way to attosecond quantum control over electronic resonance phenomena in more complex systems.

\bibliography{pop_references}
\bibliographystyle{Science}

\section*{Acknowledgments}
We thank A. Schneider and M. Seiler for their technical support. \textbf{Funding} M.H. acknowledges the funding from the European Union’s Horizon 2020 research and innovation program under the Marie Skłodowska-Curie grant agreement No 801459 - FP-RESOMUS. This work was supported by ETH Z\"urich and the Swiss National Science Foundation through projects 200021\_172946 and the NCCR-MUST.

\noindent\textbf{Authors contributions} M.H. and K.U. conceived the study. M.H. performed the experiments with the support of J.J. and L.C.S.. H.L. and M.H. analyzed and interpreted the data. Simulations were implemented by H.L.. J.M.R., K.U. and H.J.W. supervised the realization of the study. M.H., H.L., J.M.R, K.U. and H.J.W. wrote the paper with the input of all co-authors.

\noindent\textbf{Competing interests} None to declare. 

\noindent\textbf{Data and materials availability} All data needed to evaluate the conclusions in the paper are present in the paper or the supplementary materials.

\clearpage

\clearpage
\setcounter{page}{1}

\end{document}


\baselineskip20pt
\maketitle 

\newpage

\section*{Experimental details} 
The near-infrared laser pulse (2 mJ) was delivered from a regenerative Ti:sapphire laser amplifier at the central wavelength of 799 nm with a repetition rate of 5 kHz. The pulse duration was measured to be 31 fs for the full width at half maximum (FWHM) by a home-made transient-grating FROG. This laser beam was split with a 70:30 beam splitter, and the more intense part was focused by a silver mirror (f = 40 cm) onto a 4 mm long, argon-filled gas cell to generate the linearly polarized extreme-ultraviolet attosecond pulse train via HHG. The residual driving IR field was blocked by a 200nm-thick aluminum filter, and the transmitted XUV beam was focused into the main chamber of the COLTRIMS by a nickel-coated toroidal mirror (f = 50 cm). The XUV spectrum was characterized with a home-built XUV spectrometer consisting of an aberration-corrected flat-field grating (Shimadzu 1200 lines/mm) and a micro-channel-plate (MCP) detector coupled to a phosphor screen. The beam with $30\%$ energy was used as the dressing field in the RABBIT experiments. The dressing IR field is parallelly  polarized  with the XUV field and its intensity was controlled at a very low level (about $10^{12}$ W/cm$^2$) by an iris, which was characterized by the $U_p$ shift of photoelectron peaks compared with the XUV-only data. The dressing IR beam was recombined with the XUV beam by a perforated mirror and then was focused by a perforated lens (f = 50 cm). In the arm of the dressing field, there were two delay stages, i.e., a high precision direct-current motor (PI, resolution 100 nm) and a piezoelectric motor (PI, resolution 0.1 nm), operating on femtosecond and attosecond time scales, respectively. A cw HeNe laser beam was sent through the beam splitter to monitor the relative lengths of the IR paths in the two arms of the interferometer. A fast CCD camera behind the perforated recombination mirror was used to image the interference fringes from the HeNe laser in order to lock the phase delay between XUV and IR by the PID feedback. When scanning the XUV-IR phase delay with attosecond precision in the measurements, the piezo delay stage was actively stabilized with a step size of 167 as and a jitter of less than 30 as. For the COLTRIMS setup, the supersonic gas jet of helium atoms (backing pressure at 3 bar) was delivered along the $x$ direction by a small nozzle with a opening hole diameter of 30 $\mu m$ and passed through two conical skimmers (Beam Dynamics) located 10 mm and 30 mm downstream with a diameter of 0.2 mm and 1 mm, respectively. For the COLTRIMS spectrometer, a very weak static electric ($\sim$ 0.813 V/cm) and magnetic ($\sim$ 4.680769 G) fields were applied along the $z$ axis to collect the charged fragments in coincidence, in order to increase the energy resolution in the zero-energy regime. Only the single-ionization events (one electron is coincident with one He$^+$) were analyzed and presented in this work, and the photoelectron emission angle was integrated over its $4\pi$ range.

\section*{Long-range pump-probe experimental data}
\begin{figure}
\centering
\includegraphics[width=14 cm]{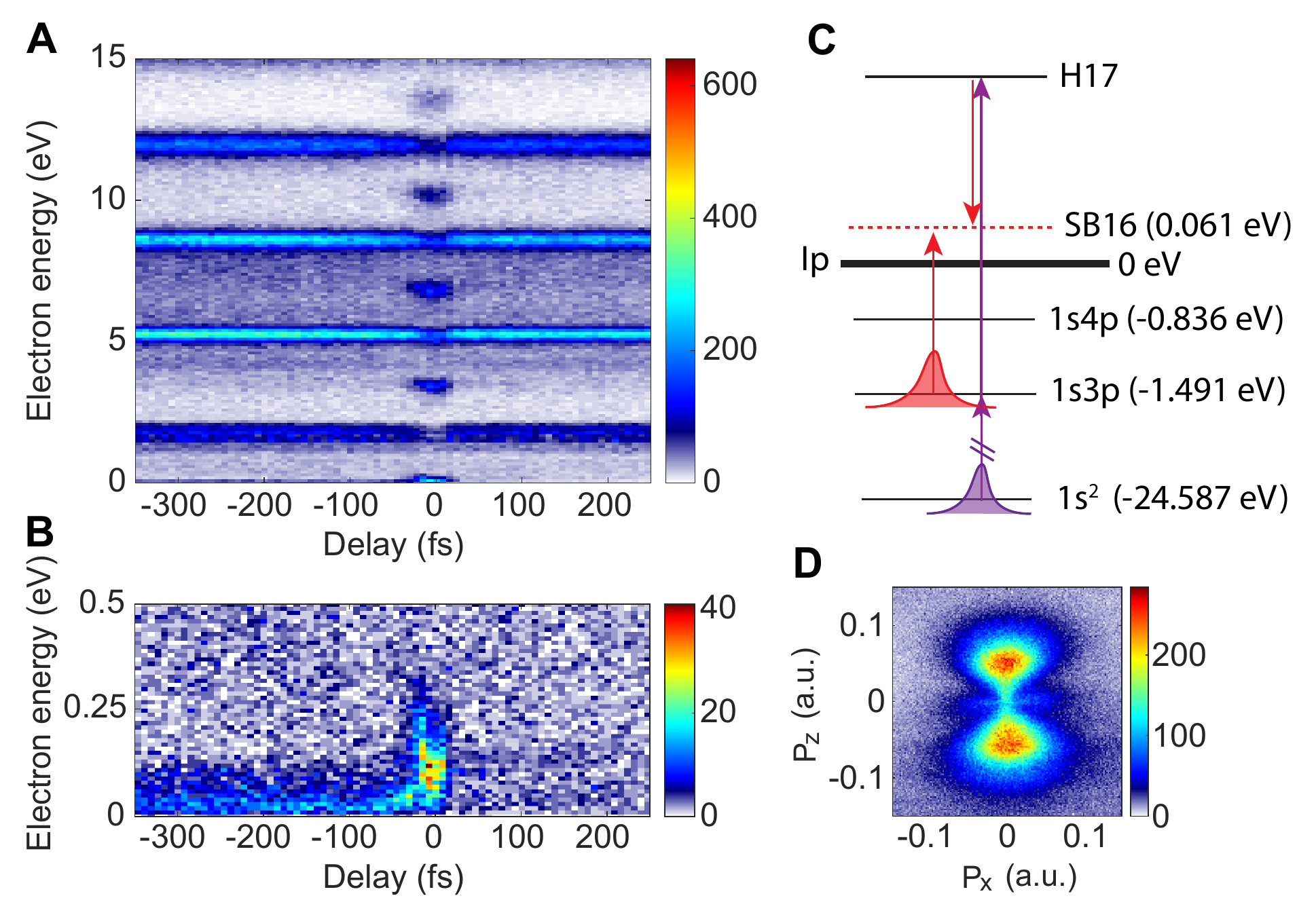}
\caption{\textbf{Long-range pump-probe experimental result}. (\textbf{A}) Measured delay-resolved photoelectron energy spectrum. The negative delay represents that the IR is behind of the XUV. (\textbf{B}) The enlarged plot of (A) for zooming in the zero-energy regime. (\textbf{C}) Schematic of the pump-probe experiment for identifying the resonant state. The reference values of the energy levels are provided by the National Institute of Standards and Technology (NIST) database. Comparing the reference values with our experimental result, we can conclude that only the 1s3p state is resonantly excited by our XUV pulse. (\textbf{D}) Photoelectron angular distribution of the probed low-energy signal recorded at the delay of -200 fs. The $d$-wave character of the probed photoelectrons indicates the XUV-pumped resonant state has the $p$-wave character. }\label{fig:long_scan}
\end{figure}

In the main text, the presented experimental data was measured with an attosecond-precision piezo stage operated in the phase-lock mode with a cw HeNe laser. Here, we display the long-range pump-probe experiment data measured with a direct-current stage motor of femtosecond precision, in order to identify the resonant state. In Fig. \ref{fig:long_scan}A, we show the measured photoelectron energy spectrum in the delay range from -350 fs to 250 fs. The common sideband signals without the resonant dynamics, such as the SB18 at 3.4 eV,   appear in the temporal overlap regime only. The SB16 with the resonant dynamics displays a long tail around zero energy beyond the overlap regime, as illustrated in Fig. \ref{fig:long_scan}B. This is a direct evidence that our XUV pulse resonantly excites the electron which is later probed by our delayed IR field. As illustrated in Fig. \ref{fig:long_scan}C, the signal  from the 1s3p state is expected at 0.061 eV, which excellently agrees with our measurement. We didn't observe any pump-probe signals around 0.716 eV, where the signal from the 1s4p state should be located, indicating our harmonic 15 only covers a single resonant state. In Fig. \ref{fig:long_scan}D, we also show the angular distribution of the probed low-energy photoelectrons at the delay of -200 fs, where the two pulses are totally separated. The $d$-wave character of the probed photoelectron indicates that the XUV-pumped resonant state has  $p$-wave character, as we expected.

\section*{Characterization of pulse durations and chirps}
\begin{figure}
\centering
\includegraphics[width=15cm]{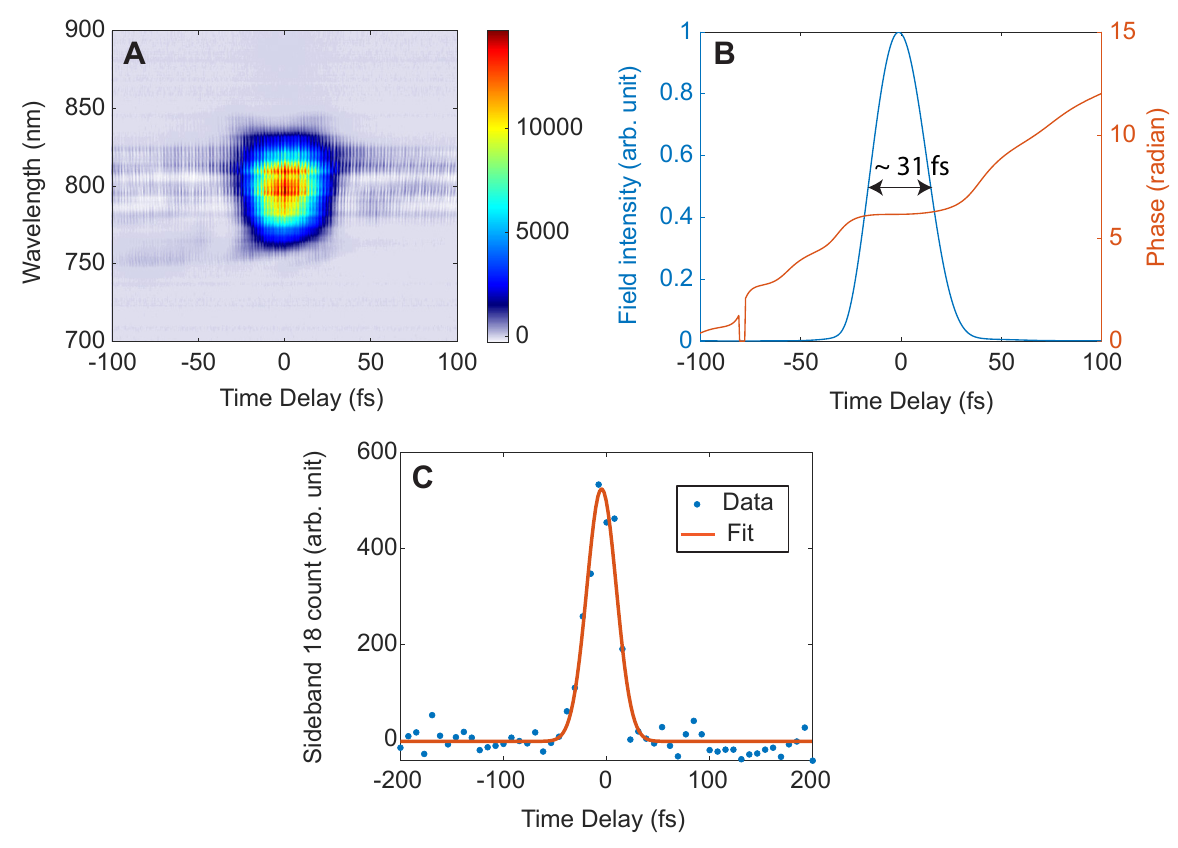}
\caption{\textbf{Characterization of IR and XUV pulses}. (\textbf{A}) Measured TG-FROG trace for the IR pulse. (\textbf{B}) Retrieved temporal envelope (blue curve) and phase (orange curve) of the IR pulse from the trace in (A). (\textbf{C}) Photoelectron signal (blue dots) of the sideband 18 and the Gaussian fit (orange curve), which is corresponding to the cross-correlation signal of the XUV and IR pulses.}\label{fig:frog}
\end{figure}

Our IR pulse was carefully characterized in a transient-gating (TG) FROG measurement, which  includes the additional dispersion of air, chamber window and lens by simulating the focus point in the COLTRIMS spectrometer chamber. In Fig. \ref{fig:frog}A, we show the raw data of the measured TG-FROG trace. Using the FROG trace, the retrieved temporal electric field and the phase are illustrated in Fig. \ref{fig:frog}B, where the full width at half maximum (FWHM) is seen to be around 31 fs. The pulse duration of the XUV field is characterized through the cross-correlation signal of the sideband 18 measured with the COLTRIMS spectrometer, as shown in Fig. \ref{fig:frog}C. The FWHM of the cross-correlation signal is measured to be around 33.3 fs determined by a Gaussian fit (orange curve in Fig. \ref{fig:frog}C). Therefore, the FWHM of the XUV pulse is estimated to be around 12 fs by the deconvolution formula assuming both pulses have Gaussian shapes. Note that we exactly used these pulse parameters in our TDSE and two-level model simulations. 

\begin{figure}
    \centering
    \includegraphics[width=10cm]{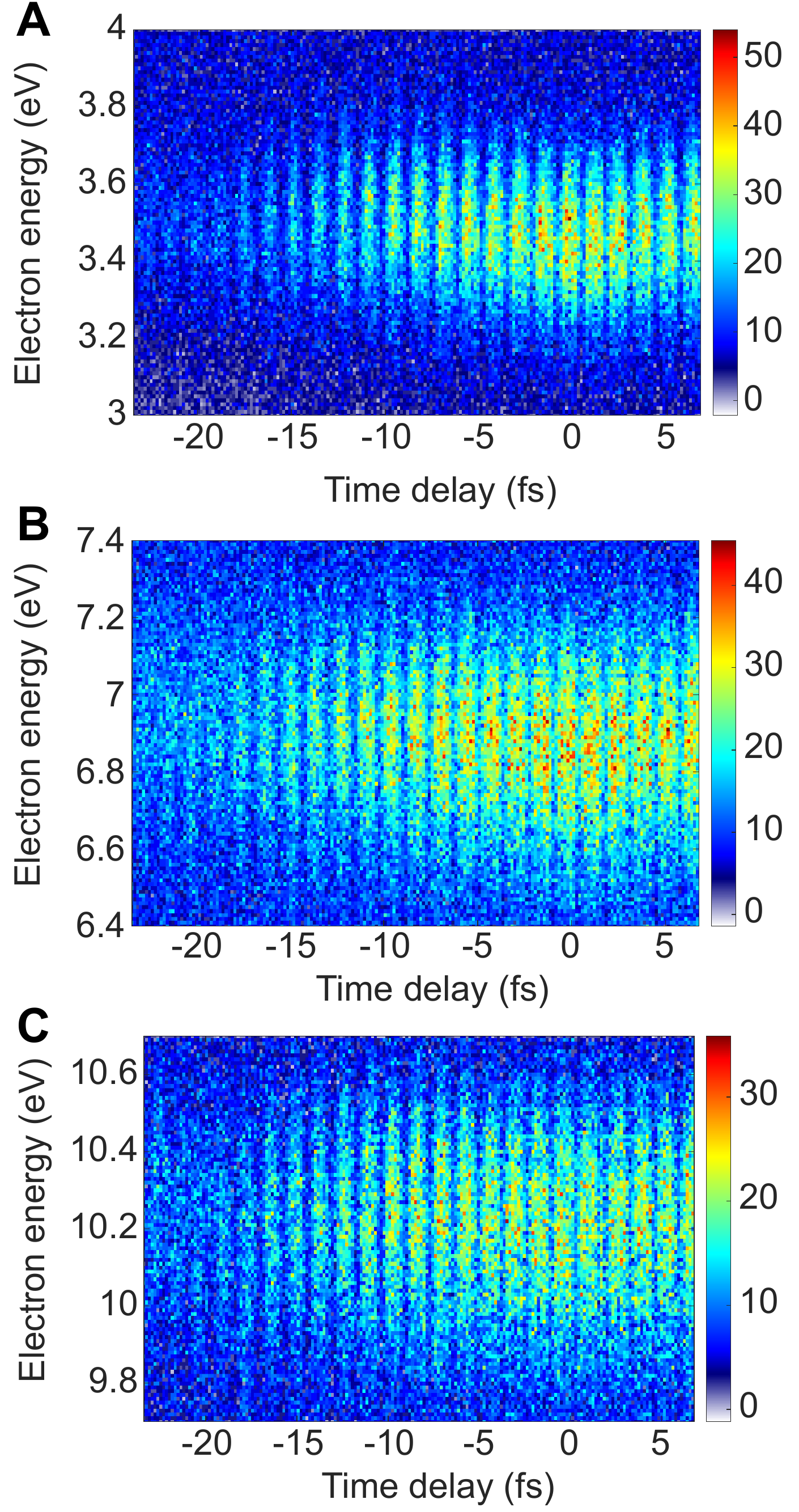}
    \caption{\textbf{Spectrally-resolved photoelectron RABBIT traces of high-order sidebands without resonant dynamics.}. (\textbf{A-C}) Measured delay-resolved photoelectron energy spectra for SB18 (A), SB20 (B), and SB22 (C). For these common sidebands, the straight RABBIT traces indicate that the femtosecond chirp of the XUV field can be neglected.}
    \label{fig:high_SB}
\end{figure}

Regarding the second-order dispersion (chirp) of our IR and XUV pulses, we can also characterize them well. For the IR pulse, the rectangle-like TG-FROG trace indicates the chirp has been mostly removed by optimizing the grating compressor. For the XUV pulse, the attochirp is the spectral phase difference between adjacent harmonics, and the femtosecond chirp is the common chirp for the spectral phase of each order harmonic which influences the XUV pulse envelope in the time domain. The attochirp has no influence for our data analysis, since we only concerned with the spectrally-resolved RABBIT phase within one order sideband. The femtosecond chirp will indeed influence the spectrally-resolved RABBIT phase within one order sideband, i.e., making the RABBIT fringes tilt. In Fig. \ref{fig:high_SB}, we show the spectrally-resolved RABBIT traces of the common sidebands without the resonant dynamics. Our experimental data shows that the spectrally-resolved RABBIT fringes are almost straight, indicating that the femtosecond chirp of our XUV pulse is very small. Therefore, what we have observed on the resonant sideband 16 and presented in the main text can be safely attributed to the resonant dynamics. Moreover, in our TDSE and two-level model simulations,  IR and XUV pulses are both chirp free. Nevertheless,  the simulated results agree well with our measurement, validating that the effect of our pulse chirps is very small. 

\newpage

\section*{TDSE simulations}

To simulate the experimental results, we solve the time-dependent Schr\"odinger equation for the Helium atom under single-active electron and dipole approximations in the velocity gauge, i.e.,
\begin{equation}
    \mathrm i\partial_t\psi = \left[\frac{\bm p^2}{2}+V(r)+p_zA(t)\right]\psi\,.
\end{equation}

The angular part of the single-electron wavefunction $\psi$ is expanded in spherical harmonics $Y_{lm}$, with $l_{\max}=3$ and $m_{\max}=0$, which is enough to cover all relevant partial waves. The radial coordinate ranging from $0$ to $100\,$a.u. is discretized into $48$ elements, where each element is represented by $8$ basis functions  based on the finite-element discrete-variable-representation~\cite{rescigno_numerical_2000}.
The model potential has the form of
\begin{equation}
    V(r) = -\frac{2+a_1\mathrm{e}^{-a_2r}+a_3r\mathrm{e}^{-a_4r}+a_5\mathrm{e}^{-a_6r}}{r},
\end{equation}
which has been adapted from Tong and Lin's work~\cite{tong05a}. The parameters have been slightly modified to match energies of all relevant bound states. In table~\ref{tab:levels} we compare the energy levels from the NIST standard, calculations with the original Tong-Lin's parameters and from the modified parameters $a_1=1.231,a_2=0.662,a_3=-2.966,a_4=2.883,a_5=-0.231,a_6=0.121$. The results with our current parameters agree better with the NIST values.
\begin{table}[h]
    \centering
    \begin{tabular}{cccc}
    \toprule
        Levels & NIST standard & Tong \& Lin & Current work \\
        \midrule
        1s$^2$ & -0.903570 & -0.903801 & -0.903540 \\ 
        1s2s & -0.145954 & -0.160010 & -0.145948 \\
        1s3p & -0.055138 & -0.056783 & -0.055135 \\
        \bottomrule
    \end{tabular}
    \caption{Energy levels of the Helium atom in atomic units from NIST standard, calculations with Tong \& Lin's model potential~\cite{tong05a} and with the current potential parameters.}
    \label{tab:levels}
\end{table}

Setting the 1s$^2$ as the initial state, we used the light pulse 
\begin{equation}
    A(t) = \sum_{i} \frac{\sqrt{I_i}}{\omega_i} \exp\left[{-4\ln 2\left(\frac{t-\tau_i}{T_i}\right)^2}\right]\cos\omega_i(t-\tau_i)\,,
\end{equation}
where $i=$IR, H15, H17 corresponds to the fundamental laser field, $15$- and $17$-th order harmonics of the fundamental field, respectively. Thus, one has $\omega_\text{IR}=0.05703\,$a.u., $\omega_\text{H15}=15\omega_\text{IR}$ and $\omega_\text{H17}=17\omega_\text{IR}$. $T_i$ is the full width at half maximum of the $i$-th pulse  with  the explicit values $T_\text{IR}=31\,\rm{fs}$ and $T_\text{H15}=T_\text{H17}=12\,\rm{fs}$. The  center $\tau_i$ of the Gaussians stands for the time delay $\tau$ between IR and XUV pulses. We set $\tau_\text{IR}=-\tau$ and $\tau_\text{H15}=\tau_\text{H17}=0$ for simplicity, which agrees with the convention in the experiment,  that negative time delay represents the IR pulse delayed with respect to the XUV pulse. The peak intensity $I_i$ of each pulses has the value of $I_\text{IR}=3\times 10^{12}\,\rm{W/cm}^2$ and $I_\text{H15}=I_\text{H17}=10^{12}\,\rm{W/cm}^2$.

 The unitary Arnoldi propagator~\cite{park_unitary_1986} is used to propagate the wavefunction in time. By projecting the wavefunction after the pulse onto the exact continuum eigen states of the model potential, one obtains the experimental observable, the photoelectron energy distribution. The wavefunction splitting technique~\cite{chelkowski_electron-nuclear_1998} is adopted to avoid reflection from the boundary. In the TDSE simulation, the photoelectron emission angle is  integrated over its $4\pi$ range as it was done in the experiment.

\section*{Two-level model}

We formulate a two-level model to include the most essential physics of the resonance dynamics. According to  second order perturbation theory, the amplitudes of the electron in the  1s$^2$ and 1s3p states are approximately given by
\begin{equation}
\begin{aligned}
    \mathrm i\dot c_{1s} &= [E_{1s}+\delta_{1s}I_\text{IR}(t)]c_{1s},\\
    \mathrm i\dot c_{3p} &= [E_{3p}+(\delta_{3p}-\mathrm i\gamma_{3p}/2)I_\text{IR}(t)]c_{3p}+\mu_{1s,3p}A_\text{H15}(t)c_{1s}.
\end{aligned}
\label{equ:model}
\end{equation}
In Eq.~\eqref{equ:model} we ignore both, the depletion and Stark-shift effects due to the XUV pulse, whichis justified since its intensity is low. Depletion $\gamma_iI_\text{IR}(t)$ and Stark shift $\delta_iI_\text{IR}(t)$ due to the IR pulse are included, which are both simply proportional to the instantaneous IR intensity $I_\text{IR}(t)$ within second-order perturbation theory and the adiabatic approximation~\cite{liang_dynamical_2020}. Note that $\delta_{1s}I_\text{IR}(t)\approx -U_p(t)$. At the peak intensity of the IR pulse, one has $\delta_{1s}I_\text{IR}=-6.6\times 10^{-3}\,\rm{a.u.}=-0.18\,\rm{eV}$, $\delta_{3p}I_\text{IR}=2.2\times 10^{-3}\,\rm{a.u.}=0.06\,\rm{eV}$,  $\gamma_{3p}I_\text{IR}=4.8\times 10^{-3}\,\rm{a.u.}=0.20\,\rm{fs}^{-1}$, and $\mu_{1s,3p}\equiv\langle 1s|p_z|3p \rangle$ stands for the dipole transition element between the two states.

With the  amplitudes from Eq.~\eqref{equ:model}  the ionization amplitudes to the sideband 16 ($E\approx E_{1s}+16\omega_\text{IR}$) read
\begin{equation}
    M_{\rm{res}}(E,s/d) = \langle E,s/d|p_z|3p\rangle\int\mathrm e^{\mathrm iEt} A_\text{IR}(t)c_{3p}(t) \,\mathrm dt,
    \label{equ:res-trans}
\end{equation}
for the resonant channel and
\begin{equation}
\begin{aligned}
    M_{\rm{con}}(E,s/d) = &\int \mathrm dE'\,\langle E,s/d|p_z|E',p\rangle\langle E',p|p_z|1s\rangle \times\\
    &\int \mathrm e^{\mathrm iEt}A_\text{IR}(t) \,\mathrm dt\int^t \mathrm e^{-\mathrm iE'(t-t')}A_\text{H17}(t')c_{1s}(t')\,\mathrm dt' \\
    \approx &\pi\langle E,s/d|p_z|E',p\rangle\langle E',p|p_z|1s\rangle\int \mathrm e^{\mathrm iEt}A_\text{IR}(t)  A_\text{H17}(t)c_{1s}(t)\,\mathrm dt.
\end{aligned}
\label{equ:cc-trans}
\end{equation}
for the continuum transition channel. In  the last line of Eq.~\eqref{equ:cc-trans} we have assumed that the product of the dipole transition elements is a slowly varying function of $E'$, and thus the integral of $E'$ is a delta-function. Note that all integrals over time without explicit integration limits in this section run from  minus infinity to plus infinity. For the results of the two-level model presented in the main text, we have solved  Eqs. (4-6) numerically. 

Next, we move to discuss the role of the Wigner delay and CC delay in our scheme. In the resonant pathway (Eq. ~\eqref{equ:res-trans}), the dipole transition moment $\langle E|p_z|3p\rangle$ gives rise to the Wigner phase at the sideband energy $E$. 
In the continuum pathway (Eq.~\eqref{equ:cc-trans}), the dipole transition moments $\langle E'|p_z|1s\rangle$ and $\langle E|p_z|E'\rangle$ contribute the Wigner phase at $E'= E + \omega$ and the CC phase connecting $E$ and $E'$, respectively. Since $\arg\langle E|p_z|E'\rangle\langle E'|p_z|1s\rangle = \arg\langle E|p_z|1s\rangle$ and $\arg\langle E|p_z|1s\rangle = \arg\langle E|p_z|3p\rangle$,  Wigner and CC phases between the two pathways  approximately cancel each other for  final states of both angular momenta, $l=0$ and $l=2$. This is confirmed by the comparison between the results from the two-level model and the TDSE. Note that for the case of normal sidebands without crossing the threshold, there are two CC phases connecting the two Wigner phases which cannot cancel, since  one CC phase is for absorption and the other one for emission. 

Now we will add Eq.~\eqref{equ:res-trans} and Eq.~\eqref{equ:cc-trans} together. Within the rotating wave approximation, one can simplifiy  the time-dependent vector potential $A_i(t)\approx \sqrt{I_i(t)}\mathrm e^{\pm\mathrm i\omega_i(t-\tau_i)}/2\omega_i$. Finally, we obtain a closed form for the total ionization amplitude
\begin{equation}
\begin{aligned}
M_{\rm{total}}(E,s/d) =& \frac{\mu_{1s,3p}\mu_{3p,E}}{4\omega_\text{IR}\omega_\text{H15}}\mathrm e^{-\mathrm i\omega_\text{IR}\tau}\int \mathrm dt\int^t \mathrm dt' \sqrt{I_\text{IR}(t)I_\text{H15}(t')}\times\\
&\exp\left\{-\mathrm i[\Phi_{3p}(t)+\omega_\text{IR}t-Et-\Phi_{3p}(t')+\Phi_{1s}(t')-\omega_\text{IR}t']\right\}+\\
    & \pi\frac{\mu_{1s,E'}\mu_{E',E}}{4\omega_\text{IR}\omega_\text{H17}}\mathrm e^{\mathrm i\omega_\text{IR}\tau}\int \sqrt{I_\text{IR}(t)I_\text{H17}(t)}\exp\left\{-\mathrm i[-\omega_\text{IR}t-Et+\omega_2t+\Phi_{1s}(t)]\right\}\,\mathrm dt.
\end{aligned}
\end{equation}
in which
\begin{align}
    \Phi_{1s}(t)&\equiv \int^t [E_{1s}+\delta_{1s}I_\text{IR}(t')]\,\mathrm dt',\\
    \Phi_{3p}(t)&\equiv \int^t [E_{3p}+(\delta_{3p}-\mathrm i\gamma_{3p}/2)I_\text{IR}(t')]\,\mathrm dt'.
\end{align}
As we have discussed previously, Wigner and CC phases cancel out each other approximately, the phase difference between $M_{\rm{res}}(E,s/d)$ and $M_{\rm{con}}(E,s/d)$ is solely given by $-2\omega_\text{IR}\tau$ in exponential and the integrals. The latter is associated with the resonant dynamics and  independent of the angular momentum $s$ or $d$ of the final state. By numerical determination of the integrals in Eq. (7), one can determine the phase difference and then calculate the interference minima ($\Delta\varphi = (2n+1)\pi$) as shown in Fig. 2B of the main text, where the dashed curves are overall shifted  by a small delay constant (0.25~fs) to match the TDSE fringes. This is the only adjustment necessary to achieve very good agreement with the experimental results.

One can further analyze the phase information of those integrals. The integral for the continuum pathway is just a Fourier transformation of the product of the pulse envelopes $\sqrt{I_\text{IR}(t)I_\text{H17}(t)}$, if one ignores the time-dependence of  $U_p$ shift. The product of two Gaussian functions is still a Gaussian function, centered at $\tau_* = -\tau/(1+T_\text{IR}^2/T_\text{H15}^2)$, corresponding to the solid green line  in Fig. 4 of the main text. The $U_p$ shift simply induces the chirps on the Gaussian wavepacket, depending on the sign of $\tau_*$. 

As for the resonant pathway, the double integral shows a memory effect: ionization at time $t$ depends on the intensity $I_\text{H15}$ at time $t'<t$. The lifetime of the memory is determined by the imaginary part of $\Phi_{3p}$. It gives an additional time delay to the resonance ionization channel, as we have seen in Fig.~4 of the main text. If one treats $I_\text{IR}(t)$ as a time-independent constant, the complex double integral becomes the Fourier transformation of a convolution
\begin{equation}
\begin{aligned}
    M_{\rm{res}}(E)&\sim \int \mathrm dt\,\mathrm e^{\mathrm i\Delta_2t}\int^t \mathrm dt'\, \sqrt{I_\text{IR}I_\text{H15}(t')}\mathrm e^{\mathrm i(\Delta_1+\mathrm i\Gamma/2-\Delta_2)(t-t')}\\
    &\sim \frac{-\mathrm i}{\Delta_1+\mathrm i\Gamma/2}\int \sqrt{I_\text{IR}I_\text{H15}(t)}\mathrm e^{-\mathrm i\Delta_2t}\,\mathrm dt,
\end{aligned}
\end{equation}
where $\Delta_1\equiv E-\omega_\text{IR}-E_{3p}-\delta_{3p}I_\text{IR}$, $\Gamma\equiv \gamma_{3p}I_\text{IR}$ and $\Delta_2\equiv E-E_{1s}-\delta_{1s}I_\text{IR}-\omega_\text{IR}-\omega_\text{H15}$ represents the detuning between 3p and final states, the decay rate of 3p state and the detuning between 1s and final states, respectively. Note that here we obtain a formula for the two-photon resonant transition with the same form as the Breit-Wigner amplitude, indicating that our physical picture based on the Fano resonance is corect. The residue Fourier transform on $\sqrt{I_\text{H15}}$ represents the suppression due to the detuning of the resonant transition.

\section*{Data analysis}

In this section, we will describe how to extract the energy- and delay-dependent interference components and the resonant electron wave packet from the original experimental pattern $P(E,\tau)$ shown in Fig.~2A of the main text.

To fit $a,b$ and $\varphi$ around the time delay $\tau$ and photoelectron energy $E$, we solve the following least-square problem
\begin{equation}
    \min_{a,b,\varphi} \sum_i|a+2b\cos(2\omega\tau_i+\varphi)-P(E,\tau_i)|^2\exp{\left[-\frac{(\tau-\tau_i)^2}{\tau_w^2}\right]},
\end{equation}
where the summation runs over all  measured $\tau_i$. The time width of the weight function $\tau_w$ is chosen to be $0.94\,\rm{fs}$, as a consequence of the trade off between time resolution and signal-to-noise ratio.

After the local cosine fitting, we extract the product of the amplitudes of two ionization channels $b=|M_\text{res}M_\text{con}^*|$ and their phase difference $\varphi=\arg(M_\text{res}M_\text{con}^*)$ as a function of electron energy $E$ and time delay $\tau$. In principle, one can divide $b\mathrm e^{\mathrm i\varphi}$ by $M_\text{con}^*$ evaluated from the two-level model to determine the resonant ionization amplitude $M_\text{res}$. However, as $M_\text{con}$ goes to zero exponentially for large $E$ and $b$ always contains some finite noises, a simple division will not produce sensible results. Rather, one has to apply a filter to reduce the contribution from the high energy part  such that the reconstructed resonant wave packet in the time domain reads
\begin{equation}
   \tilde M_\text{res}^\text{rec}(t;\tau) = \sum_{E_i} \mathrm e^{-\mathrm iE_it} b(E_i;\tau)\mathrm e^{\mathrm i\varphi(E_i;\tau)}(M_\text{con}^*)^{-1} \mathrm e^{-(E_i/E_c)^\alpha},
\end{equation}
where $\alpha=4$ and $E_c=0.34\,\rm{eV}$. We have checked that changing the filter parameters does not affect the physical signal of the ionization wave packet.

\bibliography{pop_references}
\bibliographystyle{Science}